\documentclass[prd,preprint]{revtex4}

\usepackage{graphicx}

\begin{document}

\title{The role of a dynamical measure and dynamical tension\\
in brane creation and growth}

\author{Stefano Ansoldi}
\email{ansoldi@trieste.infn.it}
\homepage{http://www-dft.ts.infn.it/~ansoldi}
\affiliation{Center for Theoretical Physics -
Laboratory for Nuclear Science\\
and Department of Physics\\
Massachusetts Institute of Technology\\
77 Massachusetts Avenue - NE25 4th floor, Cambridge, MA 02139, USA}

\author{Eduardo I. Guendelman}
\email{guendel@bgu.ac.il}
\affiliation{Department of Physics, Ben Gurion University,
Beer-Sheva 84105, Israel}

\author{Euro Spallucci}
\email{spallucci@trieste.infn.it}
\affiliation{Dipartimento di Fisica Teorica, Universit\`{a} di Trieste
and INFN, Sezione di Trieste\\
Strada Costiera, 11 -- I-34014 Miramare - Trieste (TS), Italy}



\preprint{MIT-CTP-3697}

\begin{abstract}
The use of a non-Riemannian measure of integration
in the action of strings and branes allows the possibility
of dynamical tension. In particular, lower dimensional objects
living in the string/brane can induce discontinuities in the
tension: the effect of pair creation on the string tension is
studied. We investigate then the role that
these new features can play in string and brane creation
and growth. A mechanism is studied by means of which a
scalar field can transfer its energy to the tension of
strings and branes. An infinite dimensional symmetry group
of this theory is discussed. Creation and growth of bubbles in a
formulation that requires mass generation for the bulk gauge
fields coupled to the branes is also discussed.
\end{abstract}

\maketitle

\section{Introduction}

Extended objects are presently playing an important role in theoretical
physics, as candidate theories for the unification
of all interactions, like in superstring theory
or $M$-theory \cite{bib:strbra}.
Also in cosmology, cosmic strings and domain walls have been
extensively discussed \cite{bib:1985PhRep.121....263V}.

One important question related to the dynamics of \emph{extended objects},
is whether the tension of strings and branes could be dynamical.

In \cite{bib:dynten} a framework has been discussed,
in which the tension of an extended objects is not just an \emph{external
input}, but appears as an integration constant. This formulation of the
theory of strings and branes requires the use of a non-Riemannian measure
of integration in the action for the extended object.

It is perhaps worthwhile to remark, that a non-Riemannian integration
measure has been already introduced in the context of field theories
of particles and gravitation, in order to address the problems of
the cosmological constant, of the spontaneous breaking of scale invariance,
of the fermion families, of the dark energy as well as in the brane world
scenario \cite{bib:nonrieintmea}.

In the consistent formulation of string and brane theories with dynamical
tension, internal gauge fields have also to be introduced. Then, the
integration of the equations of motion derived from the variation of these
gauge fields, gives rise to a dynamical string and brane tension.
The coupling of the gauge fields to the brane world-currents allows the tension
to become discontinuous in the case of point-like sources.

As we will discuss here, a scalar field defined in the bulk can be used
to induce an appropriate world-sheet current, that then affects directly
the tension of the extended objects it couples to.

This provides a way to directly \emph{pump} the energy, that the scalar
field has, into a network of string or branes: such a process can be
of interest in cosmology.

From the more formal point of view, it is also of interest to understand what
kind of symmetries are possessed by this kind of description of an extended
object: we will see that an infinite dimensional set of symmetries is present.

Finally we also discuss the possibility of bubble creation and growth in a
formulation where the coupling to external antisymmetric gauge fields is
modulated by the same factor that affects the string tension, especially in
connection with the necessity of mass generation for the antisymmetric gauge
fields.

\section{Strings and Branes with dynamical tension}

There is an intimate connection between the concept of strings and branes
with dynamical tension and the possibility to use a modified measure
of integration in their action.

Indeed, when performing the integration in the action functional for the
extended object, one should use an invariant volume element. The standard
(Riemannian) volume element is
\begin{equation}
    \sqrt{- \gamma} d ^{p} \sigma ,
\label{eq:rievolele}
\end{equation}
where $\gamma = \det(\gamma _{ab})$ is the determinant of the metric
$\gamma _{ab}$ defined on the manifold. Given a reparametrization of
the coordinates $\sigma ^{a}$, the volume element (\ref{eq:rievolele})
is, of course, invariant.

This invariance can be achieved also if, instead of $\sqrt{- \gamma}$, another
\emph{density} is used. For example, given $p$ scalars $\varphi _{a}$,
$a = 1 , \dots{} , p$, one can construct the density
\begin{equation}
    \Phi
    =
    \epsilon _{a _{1} \dots{} a _{p}}
    \epsilon ^{\mu _{1} \dots{} \mu _{p}}
    \partial _{\mu _{1}} \varphi ^{a _{1}}
    \dots
    \partial _{\mu _{p}} \varphi ^{a _{p}}
    ,
\label{eq:altvolele}
\end{equation}
where $\epsilon ^{\mu _{1} \dots{} \mu _{p}}$ and $\epsilon _{a _{1} \dots{} a _{p}}$
are the alternating symbols. With this definition, $\Phi$ transforms exactly
as $\sqrt{- \gamma}$ under a reparametrization transformation (which means that
$\Phi / \sqrt{- \gamma}$ is a scalar).

A straightforward use of the measure (\ref{eq:altvolele}) in string theory is
somewhat problematic however. Indeed, if in the Polyakov action \cite{bib:strbra}
\begin{equation}
    S _{\mathrm{P}} [ X ^{\alpha} , \gamma _{mn}]
    =
    - T
    \int d \sigma ^{0} d \sigma ^{1} \sqrt{- \gamma}
        \gamma ^{ab} \partial _{a} X ^{\mu} \partial _{b} X ^{\nu} g _{\mu \nu}
\label{eq:Polact}
\end{equation}
we simply replace $\sqrt{- \gamma}$ by
\begin{equation}
    \Phi
    =
    \epsilon ^{ab} \epsilon _{ij}
    \partial _{a} \varphi ^{i}
    \partial _{b} \varphi ^{j}
\label{eq:strvolele}
\end{equation}
we obtain the action
\begin{equation}
    S _{1}
    =
    -
    \int d \sigma ^{0} d \sigma ^{1} \Phi
        \gamma ^{ab} \partial _{a} X ^{\mu} \partial _{b} X ^{\nu} g _{\mu \nu}
\label{eq:strmodact}
\end{equation}
(an overall factor $T$ in (\ref{eq:strmodact}) is \emph{now irrelevant} since
it can be eliminated by simply rescaling the $\varphi ^{j}$ fields). However,
the action (\ref{eq:strmodact}) is not satisfactory, since the variation
with respect to $\gamma ^{ab}$ gives
\begin{equation}
    \Phi \partial _{a} X ^{\mu} \partial _{b} X ^{\nu} g _{\mu \nu} = 0
    ,
\label{eq:gamvarstrmodact}
\end{equation}
which means that either $\Phi = 0$ or that the induced metric on the string
vanishes.

To improve the situation and obtain a more satisfactory action, which
still uses the measure $\Phi$, we notice that the use of a measure $\Phi$
opens new possibilities for allowed contributions to the action: indeed, let us
consider, for instance, the case in which $\sqrt{- \gamma} L$ is a total derivative;
then changing the measure, it could certainly be the case that $\Phi L$ is not a
total derivative anymore: this is exactly the situation if
\begin{equation}
    L = \frac{\epsilon ^{ab}}{\sqrt{- \gamma}} F _{ab}
    ,
\label{eq:epsFabter}
\end{equation}
where $F _{ab} = \partial _{a} A _{b} - \partial _{b} A _{a}$.
So, if we consider the action
\begin{equation}
    S = S _{1} + S _{\mathrm{gauge}}
    ,
\label{eq:totmodact}
\end{equation}
where $S _{1}$ is given by (\ref{eq:strmodact}) and
\begin{equation}
    S _{\mathrm{gauge}}
    =
    \int d \sigma ^{0} d \sigma ^{1} \Phi
    \frac{\epsilon ^{ab}}{\sqrt{- \gamma}} F _{ab}
    ,
\label{eq:gaumodact}
\end{equation}
we can see that (\ref{eq:totmodact}) is now much more interesting.
First, let us note that it is conformally invariant,
provided the fields $\varphi ^{i}$ are transformed as
\begin{equation}
    \varphi ^{i}
    \quad \longrightarrow \quad
    {\varphi ^{\prime}} ^{i}  = {\varphi ^{\prime}} ^{i} (\varphi ^{j})
    , \quad
    \Phi
    \quad \longrightarrow \quad
    J \Phi
    ,
\label{eq:modmeatra}
\end{equation}
where $J$ is the Jacobian of the transformation of the $\varphi ^{i}$ fields, and
\begin{equation}
    \gamma _{ab}
    \quad \longrightarrow \quad
    \gamma _{ab} ^{\prime} = J \gamma _{ab}
    .
\label{eq:usumettra}
\end{equation}
Moreover the variation of the action (\ref{eq:totmodact}) with respect to
$\varphi ^{j}$ gives
\begin{equation}
    \epsilon ^{ab}
    \partial _{b} \varphi ^{j}
    \partial _{a}
        \left(
            - \gamma ^{cd}
            \partial _{c} X ^{\mu}
            \partial _{d} X ^{\nu}
            g _{\mu \nu}
            +
            \frac{\epsilon ^{cd}}{\sqrt{- \gamma}} F _{cd}
        \right)
    =
    0
.
\label{eq:phivarmodtotact}
\end{equation}
If $\det (\epsilon ^{ab} \partial _{b} \varphi ^{j}) \neq 0$, which is
always true if $\Phi \neq 0$, then (\ref{eq:phivarmodtotact}) means
that all the derivatives of the quantity inside the round brackets are
zero, i.e. this quantity is a constant:
\begin{equation}
    -
    \gamma ^{cd}
    \partial _{c} X ^{\mu}
    \partial _{d} X ^{\nu}
    g _{\mu \nu}
    +
    \frac{\epsilon ^{cd}}{\sqrt{- \gamma}}
    F _{cd}
    =
    M
    =
    \mathrm{const.}
    \,
    .
\label{eq:solphivarmodtotact}
\end{equation}
Considering then the variation with respect to $\gamma ^{ab}$ of
the action (\ref{eq:totmodact}), which \emph{now} is \emph{non trivial},
we obtain
\begin{equation}
    - \Phi
    \left(
        \partial _{a} X ^{\mu}
        \partial _{b} X ^{\nu}
        g _{\mu \nu}
        -
        \frac{1}{2}
        \gamma _{ab}
        \frac{\epsilon ^{cd}}{\sqrt{- \gamma}}
        F _{cd}
    \right)
    =
    0
    .
\label{eq:gamvarmodtotact}
\end{equation}
Solving for $F _{cd} / \sqrt{- \gamma}$ from (\ref{eq:solphivarmodtotact})
and inserting in (\ref{eq:gamvarmodtotact}) we get
\begin{equation}
    \partial _{a} X ^{\mu}
    \partial _{b} X ^{\nu}
    g _{\mu \nu}
    -
    \frac{1}{2}
    \gamma _{ab}
    \gamma ^{cd}
    \partial _{c} X ^{\mu}
    \partial _{d} X ^{\nu}
    g _{\mu \nu}
    -
    \frac{1}{2}
    \gamma _{ab}
    M
    =
    0
    .
\label{eq:gamvarmodtotact001}
\end{equation}
The trace of the above equation, however, gives $M = 0$, so that
(\ref{eq:gamvarmodtotact001}) is nothing but the usual equation obtained
in string theory with the standard integration measure.

If we now look at the equation of motion obtained from the variation of the
gauge field $A _{a}$, we obtain
\begin{equation}
    \epsilon ^{ab}
    \partial _{b} \left( \frac{\Phi}{\sqrt{- \gamma}} \right)
    =
    0
,
\label{eq:gauvarmodtotact}
\end{equation}
which can be integrated to obtain
\begin{equation}
    \Phi = T \sqrt{- \gamma}
    .
\label{eq:solgauvarmodtotact}
\end{equation}
The integration constant $T$ has indeed the meaning of string tension.

All the above can be straightforwardly generalized to branes. Indeed,
the relevant action for a $p$-brane is
\begin{equation}
    S = S _{p} + S _{p-\mathrm{gauge}}
    ,
\label{eq:bratotmodact}
\end{equation}
where
\begin{equation}
    S _{p}
    =
    -
    \int d ^{p+1} \sigma \Phi
        \gamma ^{a b}
        \partial _{a} X ^{\mu}
        \partial _{b} X ^{\nu}
        g _{\mu \nu}
\label{eq:bramodact}
\end{equation}
and
\begin{equation}
    S _{p-\mathrm{gauge}}
    =
    \int d ^{p+1} \sigma \Phi
        \frac{\epsilon ^{a _{1} \dots{} a _{p+1}}}{\sqrt{- \gamma}}
        \partial _{[a _{1}} A _{a _{2} \dots a _{p+1}]}
\label{eq:bragaumodact}
\end{equation}
with $\Phi$ now defined in terms of $p+1$ scalar fields as
\begin{equation}
    \Phi
    =
    \epsilon ^{a _{1} \dots a _{p+1}}
    \epsilon _{j _{1} \dots j _{p+1}}
    \partial _{a _{1}} \varphi ^{j _{1}}
    \dots
    \partial _{a _{p+1}} \varphi ^{j _{p+1}}
    ;
\label{eq:bramodmea}
\end{equation}
this \emph{kind} of $p$-brane, obtained with the modified measure $\Phi$,
is only globally scale invariant. In order to obtain conformal invariance
for $p > 1$ one should use a \emph{vector} gauge field $A _{a}$, instead
of the antisymmetric tensor field $A _{a _{1} \dots a _{p}}$ (see \cite{bib:anttenfie}).
In this paper we will study, however, the globally scale invariant case
(\ref{eq:bratotmodact}), (\ref{eq:bramodact}), (\ref{eq:bragaumodact}).
There, the variation with respect to $A _{a _{1} \dots a _{p}}$ gives
\begin{equation}
    \epsilon ^{a _{1} \dots{} a _{p+1}}
    \partial _{a _{1}}
    \left(
        \frac{\Phi}{\sqrt{- \gamma}}
    \right)
    =
    0
    ,
\label{eq:gauvarbramodtotact}
\end{equation}
which again means
\begin{equation}
    \Phi = T \sqrt{- \gamma}
    ,
\label{eq:solgauvarbramodtotact}
\end{equation}
where $T = \mathrm{const.}$ is the \emph{dynamically generated} brane tension.

The equation of motion obtained from the variation of the $\varphi ^{j}$ fields
gives (if $\Phi \neq 0$)
\begin{equation}
    -
    \gamma ^{cd}
    \partial _{c} X ^{\mu}
    \partial _{d} X ^{\nu}
    g _{\mu \nu}
    +
    \frac{\epsilon ^{a _{1} \dots{} a _{p+1}}}{\sqrt{- \gamma}}
    \partial _{[a _{1}} A _{a _{2} \dots a _{p+1}]}
    =
    M
;
\label{eq:phivarbramodtotact}
\end{equation}
as in the string case, solving for the last term on the left-hand side and
considering also the equation obtained from the variation with respect to
$\gamma _{ab}$, one obtains that
\begin{equation}
    \gamma _{ab}
    =
    \frac{1 - p}{M}
    \partial _{a} X ^{\mu}
    \partial _{b} X ^{\nu}
    g _{\mu \nu}
    .
\label{eq:solgauvarbramodtotact001}
\end{equation}
If $M < 0$ (since $p > 1$), then by rescaling of the metric $\gamma _{ab}$ one can obtain
simply that $\gamma _{ab}$ is the induced metric on the brane.

We see that the action (\ref{eq:bratotmodact}) reproduces the normal
brane dynamics: the crucial difference is that now the tension becomes
a dynamical one.

\section{Coupling of Strings and Branes to external sources}

\subsection{Strings and Branes ``cutting''}

If to the action of the brane (\ref{eq:bratotmodact}) we add a coupling
to a world-sheet current $j ^{a _{2} \dots{} a _{p+1}}$, i.e. a term
\begin{equation}
    S _{\mathrm{current}}
    =
    \int d ^{p+1} \sigma
        A _{a _{2} \dots{} a _{p+1}}
        j ^{a _{2} \dots{} a _{p+1}}
    ,
\label{eq:bracuract}
\end{equation}
then the variation of the total action with respect to $A _{a _{2} \dots{} a _{p+1}}$
gives
\begin{equation}
    \epsilon ^{a _{1} \dots{} a _{p+1}}
    \partial _{a _{1}}
    \left(
        \frac{\Phi}{\sqrt{- \gamma}}
    \right)
    =
    j ^{a _{2} \dots{} a _{p+1}}
    .
\label{eq:gauvarbracurmodtotact}
\end{equation}
We thus see indeed that, in this case, the dynamical character of the tension
becomes very much relevant.

If, for example, we consider distributional sources of the \emph{delta} function
type, we can get finite discontinuities in the tension. For simplicity let us consider
the string example, where a source of the form
\begin{equation}
    j ^{0} (\sigma) = T \delta (\sigma ^{1})
    , \quad
    j ^{1} (\sigma) = 0
\label{eq:strextcur}
\end{equation}
is considered. Then the equation
\begin{equation}
    \epsilon ^{ab}
    \partial _{b}
    \left(
        \frac{\Phi}{\sqrt{- \gamma}}
    \right)
    =
    j ^{a}
\label{eq:strextcurequ}
\end{equation}
has a solution
\begin{equation}
    \frac{\Phi}{\sqrt{- \gamma}}
    =
    T \theta (\sigma ^{1})
    ,
\label{eq:strextcursol}
\end{equation}
i.e., zero string tension for $\sigma ^{1} < 0$ and finite string
tension for $\sigma ^{1} > 0$. That is, by means of the delta
function source we have \emph{cut} the string. Point like sources
of this type were considered in \cite{bib:dynten}. 
Generalizations of this to higher dimensional branes are straightforward.

Again in the string case, we will see that a new feature is obtained
by considering a current of the form
\begin{equation}
    j ^{0} (\sigma) = 0
    , \quad
    j ^{1} (\sigma) = - T \delta (\sigma ^{0})
    ;
\label{eq:strextcur001}
\end{equation}
indeed, now equation (\ref{eq:strextcurequ}) gives
\begin{equation}
    \frac{\Phi}{\sqrt{- \gamma}}
    =
    T \theta (\sigma ^{0})
    ,
\label{eq:strextcursol001}
\end{equation}
which represents the \emph{creation} of the string tension at the instant
$\sigma ^{0} = 0$ from the action of the external source.
Notice that both the sources, (\ref{eq:strextcur}) and (\ref{eq:strextcur001}),
obey the continuity equation.

As we will see in the following subsection \ref{sec:paicre}, a source of the form
(\ref{eq:strextcur001}) can be of importance to model part of the process
of pair creation in the world-sheet.

\subsection{\label{sec:paicre}Pair creation}

A piecewise combination of sources of the type (\ref{eq:strextcur}) and
(\ref{eq:strextcur001}) can be used to represent pair creation.
\begin{figure}
\begin{center}
\fbox{\includegraphics[width=6cm]{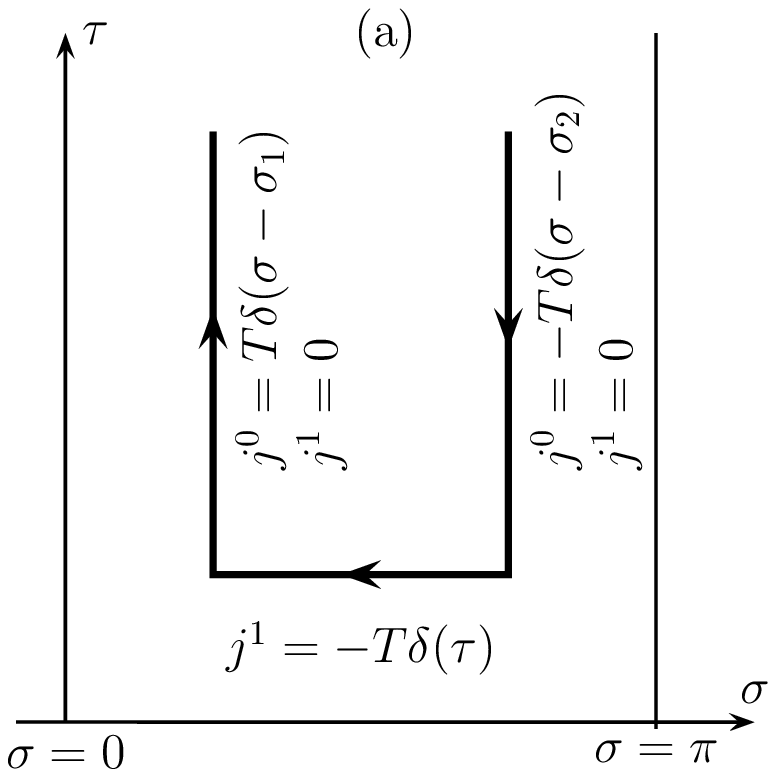}}\\
\fbox{\includegraphics[width=6cm]{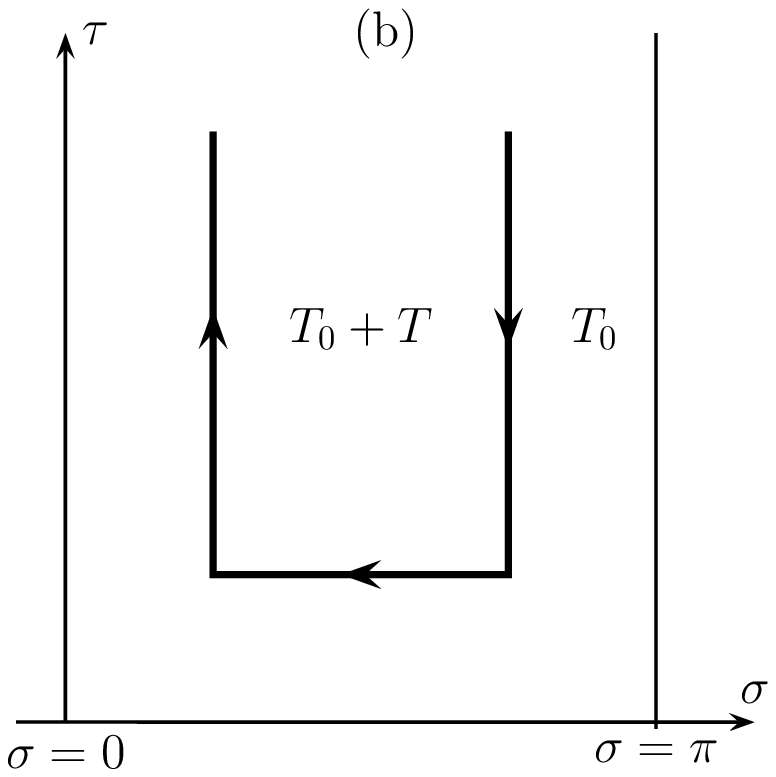}}\\
\fbox{\includegraphics[width=6cm]{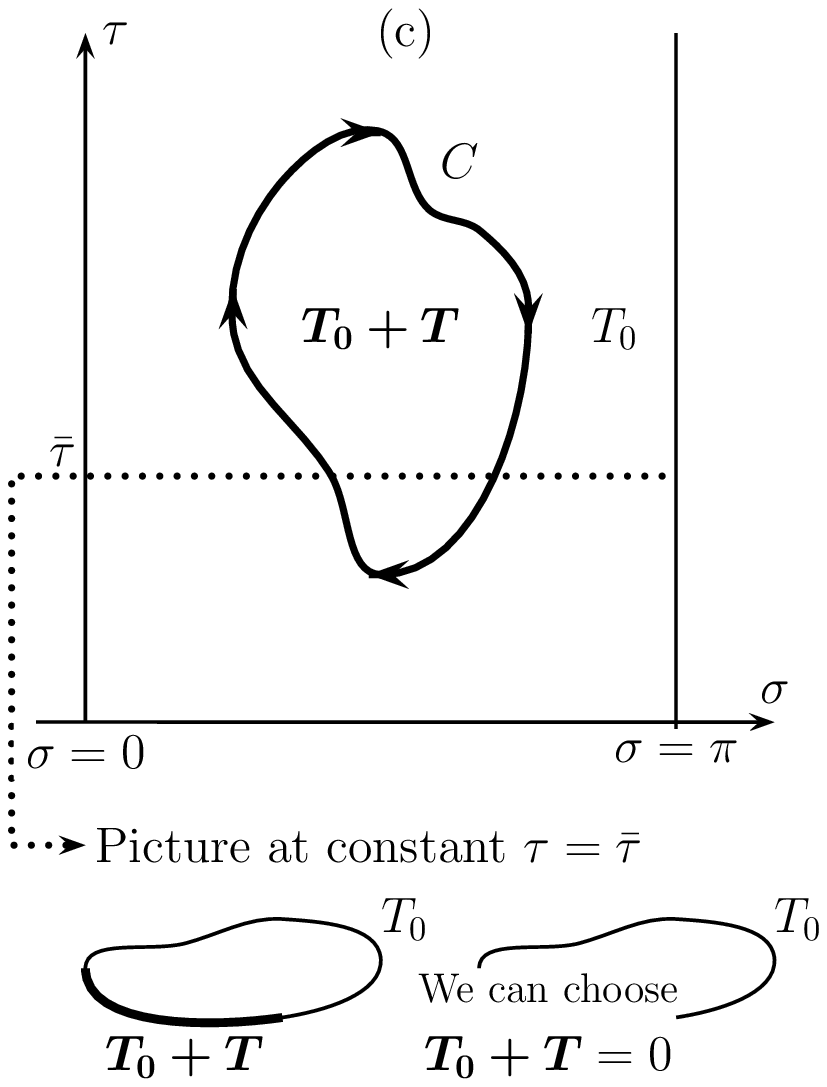}}
\end{center}
\caption{\label{fig:paicre}Some examples of models with dynamical tension,
as discussed in the main text, are presented. For example in (c) it is shown
how a string with a zero-tension segment can be obtained (please,
see the main text for details).}
\end{figure}
Take, as in figure \ref{fig:paicre}(a), a source of the form (please, note that,
for simplicity, in this subsection we are going to use $\tau$ in place
of $\sigma ^{0}$, $\sigma$ in place of $\sigma ^{1}$, so that
$\sigma ^{a} \equiv (\tau , \sigma)$, and that this
very remark also applies to figure \ref{fig:paicre}):
\begin{eqnarray}
    j ^{0}
    & = &
    T \theta (\tau) \left( \delta (\sigma - \sigma _{1}) - \delta (\sigma - \sigma _{2}) \right)
    \nonumber \\
    j ^{1}
    & = &
    - T \delta (\tau) \theta (\sigma - \sigma _{1}) \theta (\sigma _{2} - \sigma)
    ,
    \nonumber
\end{eqnarray}
with $\sigma _{2} > \sigma _{1}$.
In this case the solution of (\ref{eq:strextcurequ}) gives (see figure \ref{fig:paicre}(b))
\[
    \frac{\Phi}{\sqrt{- \gamma}} = T _{0} + T
\]
for the region $\tau > 0$, $\sigma _{1} < \sigma < \sigma _{2}$, and
\[
    \frac{\Phi}{\sqrt{- \gamma}} = T _{0}
\]
elsewhere, $T _{0}$ being an integration constant.

In the case of a pair creation and annihilation, treated in general,
the current can be described as (figure \ref{fig:paicre}(c))
\[
    j ^{a} = T \oint _{C} d \sigma ^{a} \delta ^{(2)} (\sigma ^{i} - \sigma ^{i} (\lambda))
    ,
\]
where $\sigma ^{i} = \sigma ^{i} ( \lambda )$ is a parametrization of $C$.
Then the solution of (\ref{eq:strextcurequ}) is
\[
    \frac{\Phi}{\sqrt{- \gamma}} = T _{0} + T
\]
for points inside $C$ and
\[
    \frac{\Phi}{\sqrt{- \gamma}} = T _{0}
\]
for points outside $C$, again $T _{0}$ being an integration constant.
By properly choosing the integration constant, as shown in figure \ref{fig:paicre}(c),
a string with a zero tension segment can be obtained. In more general cases,
we can also obtain strings with segments of different tension.

\subsection{Coupling to an external (scalar) field}

\subsubsection{General discussion}

Suppose that we have an external scalar field $\phi (x ^{\mu})$
defined in the bulk. From this field we can define the induced
conserved world-sheet current
\begin{equation}
    j ^{a _{1} \dots{} a _{p+1}}
    =
    g \partial _{\mu} \phi
    \frac{\partial X ^{\mu}}{\partial \sigma ^{a}}
    \epsilon ^{a a _{2} \dots a _{p+1}}
    \equiv
    g \partial _{a} \phi
    \epsilon ^{a a _{2} \dots a _{p+1}}
    ,
\label{eq:curfroscafie}
\end{equation}
where $g$ is some coupling constant.

Then (\ref{eq:gauvarbracurmodtotact}) can be integrated to obtain
\begin{equation}
    \frac{\Phi}{\sqrt{- \gamma}}
    =
    g \phi + T _{0}
    ,
\label{eq:solgauvarbracurmodtotact}
\end{equation}
which describes the transfer of energy from the bulk scalar
field $\phi$ to the brane. Notice that it is not hard to obtain from
the string point of view a current of the form (\ref{eq:strextcur001}),
which would create the tension of the string from zero tension: a scalar
field which is a step function of the time could do this.

\subsubsection{Symmetries of the system}

The part of the action which contains the gauge fields $A _{a _{2} \dots{} a _{p+1}}$,
after integration by parts, can be expressed as
\begin{equation}
    \int d ^{p+1} \sigma
        \epsilon ^{a _{1} \dots{} a _{p+1}}
        \left( \frac{\Phi}{\sqrt{- \gamma}} - g \phi \right)
        \partial _{a _{1}} A _{a _{2} \dots{} a _{p+1}}
    ,
\label{eq:gaumodbraactparint}
\end{equation}
where we remark for clarity the notation that we are using: \emph{world-sheet}
scalar fields $\varphi ^{j}$ appear in the measure $\Phi$, \emph{whereas}
the unindexed $\phi$ denotes a \emph{bulk} scalar field.
This action is of course invariant under the conventional gauge transformations
\begin{equation}
    A _{a _{2} \dots{} a _{p+1}}
    \quad \longrightarrow \quad
    A _{a _{2} \dots{} a _{p+1}}
    +
    \partial _{[ a _{2}} \Lambda _{a _{3} \dots a _{p+1}]}
    .
\label{eq:gaufiegautra}
\end{equation}
In addition, there also is the highly non conventional infinite dimensional
symmetry
\begin{equation}
    A _{a _{2} \dots{} a _{p+1}}
    \quad \longrightarrow \quad
    A _{a _{2} \dots{} a _{p+1}}
    +
    \epsilon _{a a _{2} \dots a _{p+1}}
    f ^{a} (\frac{\Phi}{\sqrt{- \gamma}} - g \phi)
    ,
\label{eq:infdimsymtra}
\end{equation}
where $\epsilon _{a a _{2} \dots a _{p+1}}$ is numerically the same as
$\epsilon ^{a a _{2} \dots a _{p+1}}$, i.e. the alternating symbol
in $p$ dimensions. In this case, using the identity
\[
    \epsilon ^{a _{1} a _{2} \dots a _{p+1}}
    \epsilon _{a a _{2} \dots a _{p+1}}
    =
    \frac{1}{p!}
    \delta ^{a _{1}} _{a}
    ,
\]
we obtain that the integrand in (\ref{eq:gaumodbraactparint}) is transformed by
\begin{equation}
    \frac{1}{p!}
    f ^{a} (\frac{\Phi}{\sqrt{- \gamma}} - g \phi)
    \partial _{a} (\frac{\Phi}{\sqrt{- \gamma}} - g \phi)
    ,
\label{eq:gaumodbraactmodter}
\end{equation}
which equals $\partial _{a} I ^{a}$, $I ^{a}$ being the integral of the
function $f ^{a}$ with respect to its argument. (\ref{eq:gaumodbraactmodter})
is therefore a total derivative and (\ref{eq:infdimsymtra}) is a symmetry.

This kind of symmetry would be absent if we would have introduced a \emph{field strength
square term}, i.e. a kinetic term for the world-sheet gauge field, and therefore
it provides a good argument for keeping linearity in
$\partial _{[a _{1}} A _{a _{2} \dots a _{p+1}]}$ in the form of the action.

One should notice that a symmetry of this kind will exist in all situations
in which one considers an antisymmetric tensor field $A _{a _{2} \dots a _{p+1}}$
coupled to a Lagrangian ${\mathcal{L}} _{1}$, which depends on dynamical
variables different from $A _{a _{2} \dots a _{p+1}}$, in the form
\begin{equation}
    S _{1}
    =
    \int d ^{p+1} \sigma
        \epsilon ^{a _{1} \dots a _{p+1}}
        \partial _{a _{1}} A _{a _{2} \dots a _{p+1}}
        {\mathcal{L}} _{1}
    .
\label{eq:gaucouL_1}
\end{equation}
Then there is the symmetry
\begin{equation}
    A _{a _{2} \dots a _{p+1}}
    \quad \longrightarrow \quad
    A _{a _{2} \dots a _{p+1}}
    +
    \epsilon _{a a _{2} \dots a _{p+1}} f ^{a} (\mathcal{L} _{1})
    .
\label{eq:gauconL_1sym}
\end{equation}
This is true whether or not (\ref{eq:gaucouL_1}) refers to a part of
the action in a $p$-brane or in a gravitational theory (i.e. it represents
a bulk gravitational action).
In this case a similar type of symmetry was discussed in \cite{bib:nonrieintmea},
where instead of the density
$\epsilon ^{a _{1} a _{2} \dots a _{p+1}} \partial _{a _{1}} A _{a _{2} \dots a _{p+1}}$
the form (\ref{eq:altvolele}) was considered. Then, given a coupling $\Phi {\mathcal{L}} _{1}$
it was shown that
\[
    \varphi ^{i}
    \quad \longrightarrow \quad
    \varphi ^{i}
    +
    f ^{i} ({\mathcal{L}} _{1})
\]
was a symmetry.

In the case of the action (\ref{eq:bratotmodact}), (\ref{eq:bramodact}),
(\ref{eq:bragaumodact}), and even when adding the coupling (\ref{eq:bracuract})
where the current does not involve the $\varphi ^{i}$ fields, there is also
the symmetry
\begin{equation}
    \phi ^{j}
    \quad \longrightarrow \quad
    \phi ^{j}
    +
    f ^{j} ({\mathcal{L}} _{1})
\label{eq:gauconL_1sym001}
\end{equation}
where
\[
    {\mathcal{L}} _{1}
    =
    -
    \gamma ^{ab}
    \partial _{a} X ^{\mu}
    \partial _{b} X ^{\nu}
    +
    \frac{\epsilon ^{a _{1} \dots a _{p+1}}}{\sqrt{- \gamma}}
    \partial _{a _{1}} A _{a _{2} \dots a _{p+1}}
    .
\]
This symmetry holds as long as there are no $\Phi ^{2}$, $\Phi ^{3}$, \dots{}
terms in the action, i.e. $\Phi$ must appear only linearly in the action.

A straightforward calculation of the Noether currents derived from the
transformation (\ref{eq:gauconL_1sym}) allows us to see that they are
\[
    j ^{a} = \frac{1}{p!} {\mathcal{L}} _{1} f ^{a} ({\mathcal{L}} _{1})
    ;
\]
indeed, since as a consequence of the equations of motion
${\mathcal{L}} _{1} = \mathrm{const.}$ we see that all components of $j ^{a}$
are constant and $\partial _{a} j ^{a} = 0$ is true on mass-shell.

\subsubsection{Discussion of possible cosmological applications}

Here we have discussed a model by mean of which it is possible to transfer energy from
a scalar field to a system of strings and/or branes and \emph{viceversa}.

Therefore very tiny strings or branes (with negligible tension) could eventually
fatten to become extended objects with greater tension. This classical process,
by means of which the energy is transferred from the scalar field
to the extended object, can replace to some extent the reheating process
in the inflationary universe: indeed, here we only need a very small amount of
energy density of strings or branes, which could then get amplified by the process.

In the context of the exchange of energy density of vacuum and dark matter (of
which the extended objects could be taken as models), we could obtain models
whereby the \emph{dark matter} and \emph{dark energy} interact.

\section{Bubble creation and mass generation}

Consider now a $2$-brane, embedded in $4$-dimensional spacetime,
which has a dynamical tension governed by
$\Phi / \sqrt{- \gamma}$. This brane has the possibility to couple
to a bulk three index antisymmetric field strength $A _{\lambda \mu \nu}$.
For example this coupling could be of the form
\begin{equation}
    e \int d ^{3} \sigma
        A _{\lambda \mu \nu} (X (\sigma))
        \partial _{a} X ^{\lambda}
        \partial _{b} X ^{\mu}
        \partial _{c} X ^{\nu}
        \epsilon ^{abc}
    ,
\label{eq:masgencou}
\end{equation}
which can be written also as
\[
    e \! \! \int \! \! d ^{4} x \! \int \! \! d ^{3} \sigma
        A _{\lambda \mu \nu} ( x )
        \partial _{a} X ^{\lambda}
        \partial _{b} X ^{\mu}
        \partial _{c} X ^{\nu}
        \epsilon ^{abc}
        \delta ^{4)} (x ^{\alpha} - X ^{\alpha }(\sigma))
\]
$e$ being some coupling constant.

This type of coupling is invariant under the gauge transformation
of the bulk field $A _{\lambda \mu \nu}$
\begin{equation}
    A _{\lambda \mu \nu}
    \quad \longrightarrow \quad
    A _{\lambda \mu \nu} + \partial _{[ \lambda} \Lambda _{\mu \nu ]}
    .
\label{eq:masgengauinv}
\end{equation}
In this case, the dynamical tension (i.e. the factor $\Phi$ present in
the other parts of the action) does not appear in the coupling of the
antisymmetric gauge field $A _{\lambda \mu \nu}$ to the world-sheet current
\[
    J ^{\alpha \beta \gamma}
    =
    e
    \int d ^{3} \sigma
        \epsilon ^{a b c}
        \partial _{a} X ^{\alpha}
        \partial _{b} X ^{\beta}
        \partial _{c} X ^{\gamma}
        \delta ^{4)} (x ^{\mu} - X ^{\mu} (\sigma))
    .
\]
If we now decide that the coupling to the external gauge field must be
part of the action which is weighted by the measure $\Phi$, we find at
first a problem with the gauge invariance (\ref{eq:masgengauinv}), since
the coupling
\begin{eqnarray}
    & &
    e
    \int d ^{4} x
    \int d ^{3} \sigma
        \frac{\Phi}{\sqrt{- \gamma}}
        \times
   \label{eq:newmeamasgencou}
   \\
    & & \qquad \times
        A _{\lambda \mu \nu}
        \partial _{a} X ^{\lambda}
        \partial _{b} X ^{\mu}
        \partial _{c} X ^{\nu}
        \epsilon ^{abc}
        \delta ^{4)} (x ^{\mu} - X ^{\mu} (\sigma))
    \nonumber
\end{eqnarray}
is not invariant anymore under (\ref{eq:masgengauinv}).

Another way to see the failure of the gauge invariance in the case
(\ref{eq:newmeamasgencou}), is by noticing that the current
\begin{equation}
    J _{\Phi} ^{\lambda \mu \nu}
    =
    e \! \!
    \int \! \! d ^{3} \sigma
        \frac{\Phi}{\sqrt{- \gamma}}
        \epsilon ^{abc}
        \partial _{a} X ^{\lambda}
        \partial _{b} X ^{\mu}
        \partial _{c} X ^{\nu}
        \delta ^{4)} (x ^{\alpha} - X ^{\alpha} (\sigma))
\label{eq:masgenmodmeacur}
\end{equation}
fails to be conserved, since it is multiplied by the dynamical
tension $\Phi / \sqrt{- \gamma}$.

If $\Phi / \sqrt{- \gamma}$ starts from zero, this represents
indeed the creation of the $2$-brane current from \emph{nothing}.
The way to restore gauge invariance, even though we couple to
a non conserved current, is to introduce the compensating
auxiliary St\"{u}ckelberg field and consider, instead of
(\ref{eq:newmeamasgencou}), the coupling
\begin{eqnarray}
    & &
    e
    \int d ^{4} x
    \int d ^{3} \sigma
        \frac{\Phi}{\sqrt{- \gamma}} \delta ^{4)} (x ^{\alpha} - X ^{\alpha} (\sigma)) \times
    \label{eq:stunewmeamasgencou}
    \\ & & \qquad \times
        \left( A _{\lambda \mu \nu} + \partial _{[\lambda } \Lambda _{\mu \nu ]} \right)
        \partial _{a} X ^{\lambda}
        \partial _{b} X ^{\mu}
        \partial _{c} X ^{\nu}
        \epsilon ^{abc}
    .
\end{eqnarray}
This is analogous to what is done in \cite{bib:2001PhReD.64..025008A}, although there only the situation
in which the brane is abruptly created at some spacelike surface was considered
(as, for instance, in some Minkowski description of a nucleation process).

In this case, the factor $\Phi / \sqrt{- \gamma}$ could be continuous
(i.e. it is not bound to be just some theta function) and represent
a continuous build up, or growth, of the $2$-brane world sheet current,
together with the brane tension.

The theory of the three index field strength coupled with a non conserved
current is not consistent if only an ordinary kinetic term proportional to
\[
    \int d ^{4} x
        \sqrt{- g}
        \left( \partial _{[\lambda} A _{\mu \nu \rho]} \right) ^{2}
\]
is considered.
Consistency is restored, however, if a mass term
\[
    m ^{2}
    \int d ^{4} x
        \sqrt{- g}
        \left( A _{\lambda \mu \nu} + \partial _{[ \lambda} A _{\mu \nu ]} \right) ^{2}
\]
is added to the action in addition to the previous kinetic term. Then
\begin{equation}
    m ^{2} \partial _{\lambda}
        \left( A ^{\lambda \mu \nu} + \partial ^{[ \lambda} \Lambda ^{\mu \nu ]} \right)
    =
    \partial _{\lambda} J _{\Phi} ^{\lambda \mu \nu}
\label{eq:divequ}
\end{equation}
follows. This is a basic relation between the divergence of
$A _{\lambda \mu \nu} + \partial _{[ \lambda} A _{\mu \nu ]}$ and the
divergence of the current $J _{\Phi} ^{\lambda \mu \nu}$. If $m ^{2} \to 0$
equation (\ref{eq:divequ}) becomes a contradiction since it forces
$J _{\Phi} ^{\lambda \mu \nu}$ to be conserved, while we know that
if $\Phi / \sqrt{- \gamma}$ has a non-trivial spacetime dependence this is not
the case. However $m \neq 0$, i.e. mass generation restores consistency.

\section{Discussion, Conclusions and Outlook}

Here we have seen what the role of dynamical tension may be in the process
of strings and branes creation. The dynamics of the string/brane tension is
governed by world-sheet currents. In particular these world-sheet currents
could be generated by lower dimensional objects living in the extended
objects, for example point particles in a string.

In the case of a string we have studied the discontinuities in the string
tension produced by these point particles and the effect induced on the
string tension by the process of pair creation taking place in the string
world-sheet.

We have also studied effects by means of which energy may be transferred
from the bulk to strings and branes. This is possible because, for example,
a bulk scalar field induces naturally a world-sheet current that affects
the tension.

A process of this kind could be of interest in the braneworld scenario.,
where energy that is transferred to the bulk could appear as ``missing``
from our $4$-dimensional world.

Other applications could be in the field of cosmology, where the classical
problem of reheating in the inflationary scenario involves the transfer
of energy from a scalar field to matter. In our model, this is realized
by the transfer of energy from a bulk scalar field to extended objects,
by altering their tension. We could thus have a situation in which a string
or brane with, initially, a very tiny tension, could grow very much by
means of the above mechanism.

It is possible to consider the action of a $2$-brane weighted by he measure $\Phi$,
including the coupling to external bulk fields $A _{\lambda \mu \nu}$. If this
is done in the way that have we described in this paper, it is also possible to
consider the creation or ``growth'' of the current coupled to $A _{\lambda \mu \nu}$.
In this case the current is \emph{not} conserved. Restoration of the gauge invariance
$A _{\lambda \mu \nu} \to A _{\lambda \mu \nu} + \partial _{[\lambda} \Lambda _{\mu \nu]}$
by considering St\"{u}ckelberg compensating fields is possible and gives rise to
a consistent model if we include mass generation for the $A _{\lambda \mu \nu}$ fields.

\begin{acknowledgments}
This work is supported in part by funds provided by the U.S. Department of
Energy (D.O.E.) under cooperative research agreement \#{}DF-FC02-94ER40818.\\
The work of S. Ansoldi is supported in part by a grant from the Fulbright Commission.\\
E. I. Guendelman would like to thank the Universities of Udine and Trieste
and I.N.F.N. and I.C.R.A. for support and hospitality.
\end{acknowledgments}

\end{document}